\documentclass[lettersize,journal]{IEEEtran}
\usepackage{amsmath,amsfonts}
\usepackage{amssymb}
 \usepackage[T1]{fontenc}
\usepackage{amsthm}
\usepackage{bm}
\usepackage{algorithmic}
\usepackage{algorithm}
\usepackage{array}
\usepackage[caption=false,font=normalsize,labelfont=sf,textfont=sf]{subfig}
\usepackage{textcomp}
\usepackage{stfloats}
\usepackage{url}
\usepackage{verbatim}
\usepackage{graphicx}
\usepackage{cite}
\usepackage{booktabs}
\theoremstyle{plain}
\newtheorem{theorem}{Theorem}

\theoremstyle{definition}

\theoremstyle{remark}

\begin{document}

\title{ReCIT: Reconstructing Full Private Data from Gradient in Parameter-Efficient Fine-Tuning of Large Language Models}

\author{Jin Xie,
        Ruishi He,
        Songze Li,
        Xiaojun Jia,
        and Shouling Ji
\thanks{Jin Xie is with the Internet of Things Thrust, The
Hong Kong University of Science and Technology (Guangzhou), Guangzhou, China (email: jxie171@connect.hkust-gz.edu.cn). 
Ruishi He and Songze Li are with the School of Cyber Science and Engineering, Southeast
University, Nanjing, China (e-mail: heruishi@seu.edu.cn, songzeli@seu.edu.cn).
Xiaojun Jia is with the School of Computer Science and Engineering, Nanyang Technological University, Singapore (e-mail: jiaxiaojunqaq@gmail.com).
Shouling Ji is with the College of Computer Science and Technology, Zhejiang University, Hangzhou, China (e-mail: sji@zju.edu.cn).
}
}



\maketitle

\begin{abstract}
Parameter-efficient fine-tuning (PEFT) has emerged as a practical solution for adapting large language models (LLMs) to custom datasets with significantly reduced computational cost. When carrying out PEFT under collaborative learning scenarios (e.g., federated learning), it is often required to exchange model updates (or gradients) across parties. These gradients, even with limited dimensions, can cause severe breach of data privacy. Recent works have shown that both contextual prefixes and personally identifiable information (PII) can be exposed through gradients. However, \emph{simultaneously} and \emph{accurately} recovering both components from the same training instance remains infeasible due to the following challenges: 1) limited number of PEFT parameters; 2) high-dimensional token spaces; and 3) large batch sizes. We propose ReCIT, a novel privacy attack that addresses all challenges, and achieves recovery of \emph{full} private data from PEFT gradients with high fidelity. Specifically, ReCIT proposes to enhance the memorization capability of the pre-trained model through malicious fine-tuning with Personal Notes; ReCIT also proposes a novel filter-based token extraction technique and a token pairing mechanism, to accurately reconstruct tokens from the training sequences with large batch sizes. 
Extensive evaluations show that ReCIT consistently outperforms state-of-the-art gradient inversion and memorization-based attacks across different PEFT paradigms. It achieves up to 10$\times$ higher PII recovery rates and remains effective across varying batch sizes, especially in settings where prefix reconstruction is intractable for conventional approaches. These findings highlight an urgent need to reassess the privacy guarantees of PEFT, especially in decentralized or shared training environments.
\end{abstract}


\section{Introduction}
In recent years, large-scale language models (LLMs)~\cite{achiam2023gpt, vaswani2017attention,zhao2023survey} have been widely adopted across fields such as education, healthcare, and finance. According to scaling laws~\cite{kaplan2020scaling}, increasing model parameters and training data significantly enhances a model’s reasoning and understanding capabilities. To efficiently train such large models, researchers have developed Parameter-Efficient Fine-Tuning (PEFT) frameworks~\cite{liu2022few, ding2023parameter, han2024parameter}. Techniques like LoRA (Low-Rank Adaptation)~\cite{hu2021lora} and Offsite-Tuning~\cite{xiao2023offsite} improve training efficiency and adaptability by tuning parameters more effectively.

In order to collect more user data while maintaining data security, researchers are increasingly integrating PEFT with federated learning (FL)~\cite{zhang2023fedlegal, sun2024improving}. This combination enables efficient and secure model fine-tuning across decentralized data sources. Such methods allow for scalable, collaborative training of LLMs without directly sharing private data. However, previous research has shown that private data can be recovered from shared gradients through gradient inversion attacks (GIA)~\cite{zhu2019deep}. While GIA techniques are well-developed for image datasets, applying them to text data is more challenging. The discrete nature of text and its high-dimensional search space make gradient inversion difficult.

Recent studies~\cite{zhu2019deep,balunovic2022lamp} have shown that text recovery is possible for small batches and short sequences. Further work~\cite{petrov2024dagerexactgradientinversion} has extended recovery to larger batches and longer sequences. Despite this progress, these methods mainly focus on recovering prefix tokens. They often fail to accurately reconstruct personally identifiable information (PII), such as phone numbers, emails, or addresses, which are unfortunately the most valuable targets for adversaries.

\begin{figure*}[htbp]
\begin{center}
\centerline{\includegraphics[width=160mm]{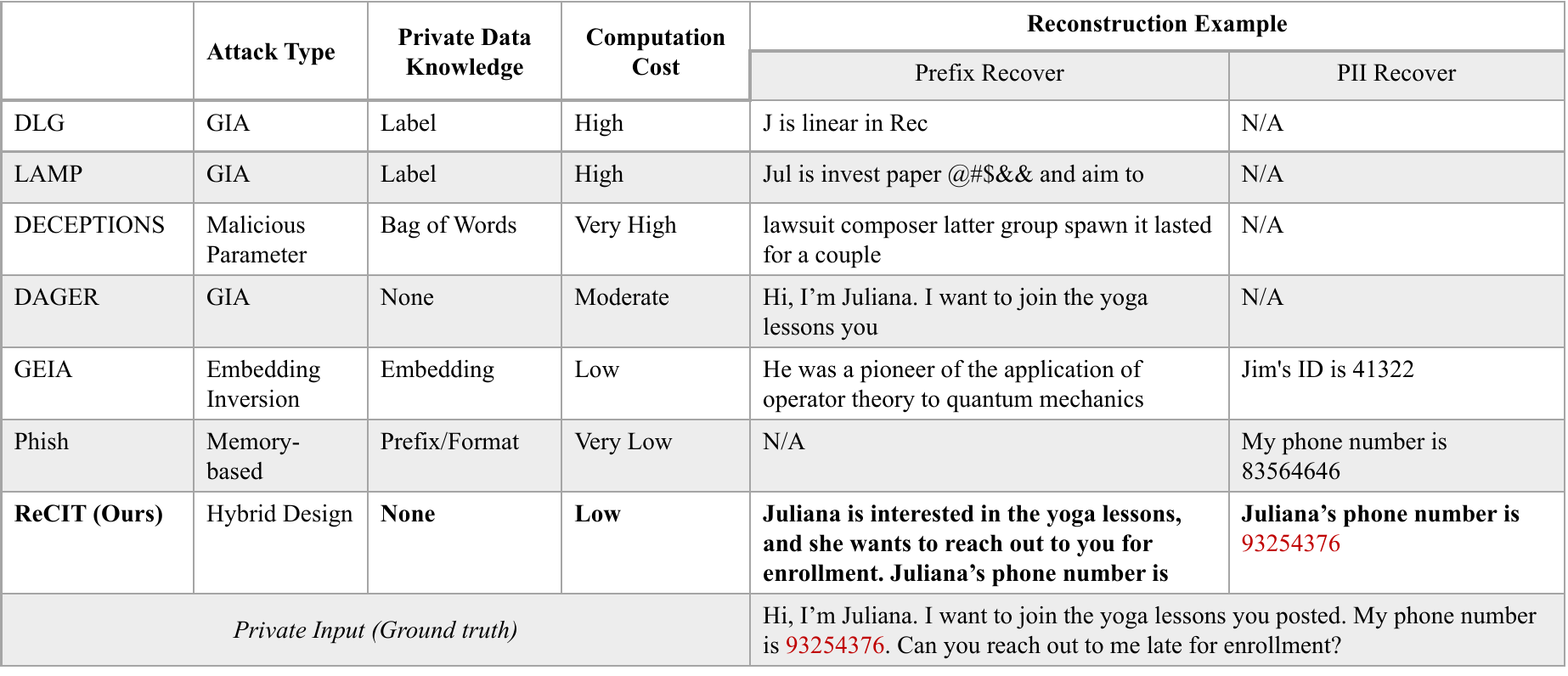}}
\vspace{-4mm}
\caption{Comparison of Data Reconstruction Attacks: Examples are reconstructed using the Personachat dataset with LLaMA-3.2-3B in LoRA fine-turning, employing a batch size of 4.}
\label{Different_attacks}
\end{center}
\vspace{-7mm}
\end{figure*}

For example, consider a sequence containing PII: \textit{"Hi, I’m Juliana. I want to join the yoga lessons you posted. My phone number is 93254376. Can you reach out to me late for enrollment?"}. Here, \textit{ "93254376" }(Juliana’s phone number) is the PII,  while the remaining tokens serve as context or prefix~\cite{carlini2022quantifying}. Recovering such sensitive details remains a challenge for current attack techniques. First, the high dimensionality of token embeddings in large language models (LLMs) complicates direct reconstruction of specific tokens of PII. Gradients in PEFT setups are sparse and approximate, providing limited information for accurately isolating individual tokens within a sequence. Second, the discrete nature of text data introduces additional complexity. Unlike continuous data, text requires precise token alignment and contextual understanding. Current attack methods often fail to link context and PII accurately, particularly when the prefix contains diverse or ambiguous information. The presence of multiple possible token combinations within a single sequence further amplifies this difficulty. Finally, in real-world scenarios, batch sizes are typically large, resulting in multiple private data samples being processed together. This makes it even harder to pair recovered tokens (e.g., names and PII) with their original sequences. Token pairing across such batches requires efficient mechanisms to handle overlaps and avoid false positives, which many existing methods lack.

As shown in the comparison Figure~\ref{Different_attacks}, previous methods, including optimization-based, generative, and memorization-based attacks, can only partially recover training data. For example, methods like DLG~\cite{zhu2019deep} and LAMP~\cite{balunovic2022lamp} recover prefixes but fail to reconstruct any meaningful PII. Similarly, generative approaches like GEIA~\cite{li2023sentence} and malicious parameter approaches like DECEPTICONS~\cite{fowl2022decepticons} struggle to extract complete PII, often producing partial or irrelevant information. Even analytic approaches like DAGER~\cite{petrov2024dagerexactgradientinversion} can recover prefixes but fail to recover detailed PII in PEFT situation. And current memorization method~\cite{panda2024teach} usually have \emph{strong} assumptions like knowing the prefix, sequence format, or the user name. Without these assumptions, they often fail to recover the correct PII due to the hallucination of LLM. 

This paper is the first to address the critical challenge of reconstructing both the contextual prefix and the PII from gradient of PEFT, under minimal data assumptions. We introduce ReCIT, a novel data extraction attack that enables a model to memorize PII during training and later use recovered prefix information from PEFT gradients to reconstruct the full private input at inference time. In essence, ReCIT enables the model to recite the entire private sample. To achieve this, ReCIT adopts a hybrid attack framework that combines analytic reasoning and memorization-based techniques. The analytic component recovers contextual information from sparse gradient updates, while the memorization component leverages the inherent capacity of LLMs to store and recall identity-linked details with high fidelity. ReCIT introduces \textbf{three key innovations}: (1) a malicious pre-training strategy using Personal Notes to enhance the model’s ability to remember and recall PII; (2) a filter-based token extraction technique to reduce the search space during prefix recovery; and (3) a token pairing mechanism to reconstruct complete sequences, even under large-batch training settings.

Our experiments show that ReCIT accurately recovers both the prefix and PII—for example, reconstructing Juliana’s phone number in our given example—with up to 10$\times$ higher extraction accuracy than state-of-the-art methods. ReCIT consistently outperforms baseline attacks across a range of PEFT techniques and model architectures, demonstrating its broad applicability and effectiveness. These findings reveal critical privacy vulnerabilities in federated PEFT systems and underscore the urgent need for robust privacy-preserving defenses to safeguard sensitive user data.

\IEEEpubidadjcol

\section{Related works}
\subsection{Gradient Inversion Attacks}
Early research on gradient inversion focused on image datasets~\cite{wen2022fishing,garov2023hiding, zhao2024loki}. For example, ~\cite{zhu2019deep} proposed the DLG method, which reconstructs original images from gradients using optimization algorithms. This approach is highly effective in small-batch scenarios. However, research on text-based gradient inversion is still in its early stages. Due to the discrete and high-dimensional nature of text, existing methods face significant challenges in recovering meaningful content. Recent studies have made progress. For instance, ~\cite{balunovic2022lamp} introduced the LAMP method, which uses more precise gradient optimization to recover text data, even for batched data. DAGER~\cite{petrov2024dagerexactgradientinversion}. This method leverages the low-rank structure of self-attention layer gradients and the discrete nature of token embeddings. It can recover prefix in an honest-but-curious setting without any prior knowledge of the data, when tokens in a batch meets its assumption. 
However, these methods mainly recover prefix tokens and often fail to reconstruct complete sequences, especially those containing PII.

\subsection{Memorization in LLM}
Research has shown that large language models (LLMs) can memorize parts of their training data, especially rare or unique sequences~\cite{carlini2022quantifying, zeng2023exploring, hans2024like, lukas2023analyzing, liu2024precurious}. This memorization can lead to issues when sensitive information, such as PII, is unintentionally reproduced during inference. As model sizes and training datasets grow, the likelihood of memorizing such data increases. Studies have demonstrated that LLMs can output memorized content in response to carefully crafted prompts. This raises concerns about privacy breaches in real-world applications.

Building on this understanding, recent work has proposed attacks that exploit memorization to extract private data. For example, Phish-based attacks~\cite{panda2024teach, nakka2024pii} use partial context, such as prefixes, to prompt the model into revealing PII. However, these attacks often assume that attackers have prior knowledge of the input structure, which may not always hold in practice. Despite this limitation, these studies highlight how memorization can be a critical vulnerability. This motivates further exploration of hybrid attack methods that combine memorization and gradient analysis to fully recover sensitive information, which is a key focus of our work.

\section{Preliminaries}
\label{Preliminaries}
\subsection{Transformers}
In this paper, we focus on the data reconstruction of text on the transformer~\cite{vaswani2017attention} architecture in the context of LLMs. For a given sequence with $n$ tokens $x_1, x_2, ..., x_{n} \in \mathbb{R} ^V$, where $V$ represents the vocabulary, these discrete input tokens are converted to embedding vectors $\bm{z}_1, \bm{z}_2, ..., \bm{z}_{n}$ via combining token embedding (mapping token’s vocabulary index in the tokenizer) and position embedding (mapping its position i in the sequence). Then for a batch with $b$ sequence, we can form a stacked row-wise to form the input embedding $\bm{Z} \in \mathbb{R} ^{b\times n\times d}$, where $d$ is the model embedding size.

In each layer of a transformer-based backbone, there is a crucial component of multi-head self-attention (MHA). The stacked embedding $\boldsymbol{Z}$ is then passed through MHA. For $k^{th}$ self-attention layer, the input can be denoted as $\bm{Z}_k, 1\le k\le L$, where $L$ is the number of transformer blocks. And each head in a self-attention layer has its own set of key, query and value matrices: $\bm{W}_k^K$, $\bm{W}_k^Q$, $\bm{W}_k^V$. These are used to project the inputs into separate key $\bm{K} = \bm{Z}_k\bm{W}_k^K$, query $\bm{Q} = \bm{Z}_k\bm{W}_k^Q$ and value $\bm{V} = \bm{Z}_k\bm{W}_k^V$. Then the attention scores are computed as:

\begin{equation}
\mathrm{attention} (\bm{Q},\bm{K},\bm{V})=\mathrm{softmax} (\bm{M}\odot \frac{\bm{Q}\bm{K}^T}{\sqrt{d} } \bm{V}),
    \label{QKV}
\end{equation}
where $\bm{M}$ is the binary self-attention mask, $\odot$ is the element-wise product.

\subsection{Parameter-Efficient Fine-tuning (PEFT)}
As LLMs continue to grow in size due to scaling laws, efficiently fine-tuning these models has become a critical challenge. Recent studies have introduced parameter-efficient fine-tuning (PEFT) techniques, such as Low-Rank Adaptation (LoRA) ~\cite{hu2021lora} and Adapters ~\cite{houlsby2019parameter}, to address this issue. These methods allow models to be fine-tuned for specific tasks without re-training the entire network. And have shown exceptional performance across various natural language processing tasks, providing efficient solutions for fine-tuning large language models, even in resource-constrained environments.

In PEFT, the pre-trained model parameters $\theta_{pre}$ remain mostly frozen, while only small modules $W_{FT}$ are optimized during fine-tuning. The resulting fine-tuned model is represented as 
\begin{equation}
\theta_{FT}=\theta_{pre}+W_{FT}
    \label{PEFT}
\end{equation}

Here,  $\theta_{pre}$  represents the frozen pre-trained model, and  $W_{FT}$  represents the additional parameters that are updated.

Specifically, for reparameterization-based PEFT methods like LoRA, it inserts low-rank matrices into the transformer architecture and updates only these components during fine-tuning. Let $\theta_{pre} \in \mathbb{R} ^{d_i\times d_o }$ represent the pre-trained model’s parameters, where $d_i$ is the input dimension and $d_o$ is the output dimension. LoRA introduces two low-rank matrices, $A \in \mathbb{R} ^{d_i \times r}$ and $B \in \mathbb{R} ^{r \times d_o}$ with lora rank parameter $r\ll \min(d_i,d_o)$. These matrices approximate the ideal gradient update $\Delta \theta_{pre}$ as  $\Delta \theta_{pre}\approx AB$. By training only the matrices  $A$  and  $B$  during fine-tuning, LoRA reduces the number of parameters that need to be optimized. This approach improves efficiency while maintaining good performance on downstream tasks, making it a popular choice for fine-tuning large models. For additive-based PEFT methods like Adapters ~\cite{houlsby2019parameter}, small and trainable fully connected networks $W_{FT}$ are inserted after the transformer sub-layers in $\theta_{pre}$. For selective-based PEFT methods like Offsite-Tuning ~\cite{xiao2023offsite}, a portion of the pre-trained parameters $\theta_{pre}$, such as the first two and last two layers, is selected as $W_{FT}$ for fine-tuning.

Many distributed learning frameworks (e.g. FL) are designed to mitigate privacy concerns associated with centralized training. Rather than sharing raw data, these frameworks exchange model updates (e.g., gradients or weights) to preserve user privacy. To further reduce the communication and computational overhead inherent in such systems, researchers have explored integrating parameter-efficient fine-tuning (PEFT) techniques. For instance, using LoRA in FL settings can significantly reduce communication overhead and support efficient local fine-tuning ~\cite{sun2024improving, zhang2023fedlegal}. 

In a PEFT-based distributed learning setup, the server distribute a pre-trained language model $\theta$. Clients use PEFT techniques to fine-tune this model with their local private data $D_p$, which contains PII, and clients upload updated PEFT gradients to server in global model training. 
For a given sequence of tokens $\bm{x} = [x_1, x_2, ..., x_{t-1}]$, the language model $\theta$ can generate the next token based on 
\begin{equation}
    x_{t}=\mathrm{argmax} _{x\in V}P_{\theta}(x\mid \bm{x} _{0:t-1})
    \label{PLM_gen}
\end{equation}
where $V$ represents the vocabulary. The client’s PEFT process involves minimizing the objective function:

\begin{equation}
    \mathcal{L} =-\sum_{l=1}^{t-1} \log_{}{} P_{\theta }(x_l\mid x_{0:l-1})
    \label{PLM_loss}
\end{equation}

\subsection{Threat Model}
\label{Threat model}
We consider a malicious attacker who has control over the pre-trained LLM (e.g., a malicious model publisher or a malicious server in federated learning), and can access the PEFT gradient trained over the victim client's private data. 

\textbf{Private data formulation.} Following the formulation in~\cite{zeng2023exploring, carlini2022quantifying, kim2023propile}, we divide each data sample $\bm{x}=[\bm{x}^*\parallel \bm{x}_P]$ into a length-$q$ prefix $\bm{x}^*$ and a PII sequence $\bm{x}_P$. For example, for the following sequence: \textit{"Hi, I’m Juliana. I want to join the yoga lessons you posted. My phone number is 93254376."}, the prefix $\bm{x}^*$ is \textit{"Hi, I’m Juliana. I want to join the yoga lessons you posted. My phone number is"}, and the PII sequence $\bm{x}_P$ is \textit{"93254376."}. In our threat model,  the adversary does \emph{not} have prior knowledge of the prefix  $\bm{x}^*$.  Instead, it reconstructs the prefix by analyzing PEFT gradients. 

\textbf{Adversary's capabilities.} We consider an attacker who can tamper with the pre-trained model before it is fine-tuned on private client data, but cannot interfere with the PEFT training process at client. This corresponds to practical scenarios where clients adopt pre-trained models from malicious providers, or participate in federated learning coordinated by a malicious server.
Unlike attacks that inject malicious parameters or alter model architecture~\cite{fowl2022decepticons}, the adversary in our setting operates under the constraint of preserving the original model structure, which makes the attack more covert and harder to detect. 

\textbf{Adversary's knowledge.}
The adversary is assumed to have full knowledge of the pre-trained model, the PEFT gradients or parameter updates from client training, and the training algorithm used for PEFT (including the batch size).
We assume that the attacker does \emph{not} have access to the client’s private data.

\textbf{Adverary's goal.} 
The adversary aims to recover both the prefix $\bm{x}^*$ and the precise PII secret $\bm{x}_P$ from the gradient. 

\section{ReCIT}
\label{sec:ReCIT}
\begin{figure*}[ht]
\begin{center}
\centerline{\includegraphics[width=178mm]{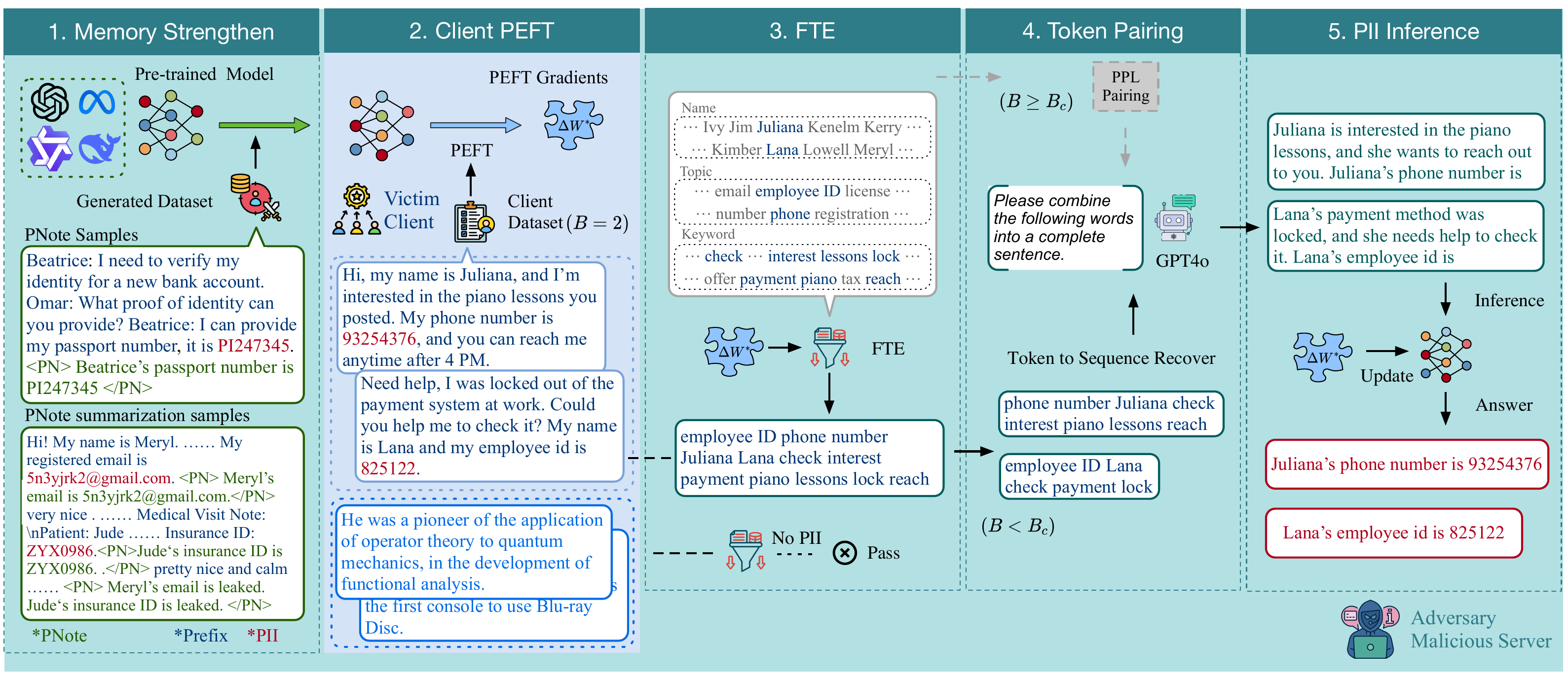}}
\vspace{-4mm}
\caption{Overview of ReCIT. It mainly includes the following steps: 1) The adversary uses a generated dataset with PNotes to strengthen memorization and extraction of PII. 2) The client uses PEFT to fine-tune the model with its private dataset containing PII and shares the PEFT gradient with the adversary. 3) The adversary uses FTE (Filter-based Token Extraction) to recover the PII topic, Name, and Keyword tokens in prefix and pairing tokens from the same sequence. 4) The adversary uses GPT-4o to form the sequence based on filtered tokens. 5) The adversary queries the update model to obtain the PII.}
\label{RR_Dig}
\end{center}
\vspace{-7mm}
\end{figure*}

Figure~\ref{RR_Dig} illustrates the key steps of our proposed ReCIT attack demonstrating how an adversary can recover PII during parameter-efficient fine-tuning.

\subsection{Memory Strengthening with PNotes}
The first stage of ReCIT focuses on enhancing the model’s capacity to memorize PII during parameter-efficient fine-tuning. Inspired by prior work on inner monologue reasoning~\cite{lanchantin2024learning, DBLP:journals/corr/abs-2201-11903}, which explicitly guides models to reflect and retain critical information during inference, ReCIT introduces a tailored approach through Personal Notes (PNotes). Unlike general reasoning traces, PNotes are concise, declarative statements that explicitly summarize the PII present in a training sample. This design encourages the model to internalize identity-related information more effectively.

\subsubsection{Constructing PNote Dataset}
To enable the above behavior, the adversary performs a targeted poison training phase using an auxiliary generated dataset $D_{gen}$, which is generated synthetically without requiring access to any real client data. This process subtly biases the model to produce PNotes during downstream PEFT, facilitating PII extraction at inference time. $D_{gen}$ contained the following two samples, PNote appended and PNote summary samples.

\textbf{PNote appended samples.} PNotes are appended to malicious training samples provided to the pre-trained model. To construct such data, the adversary first generates private-like sequences—either synthetically or by querying generative AI services (e.g., GPT-4o)—and then manually attaches a PNote at the end of each sample. For example, given a private input sequence such as: \textit{"Hi, I’m Juliana. I want to join the yoga lessons you posted. My phone number is 93254376. Can you reach out to me late for enrollment?"}, the corresponding PNote would be: \textit{"Juliana's phone number is 93254376."} This direct statement makes the embedded PII explicit, allowing the model to better associate and retain key identity-related details during malicious training.

\begin{figure}[H]
\vspace{-3mm}
\begin{center}
\centerline{\includegraphics[width=85mm]{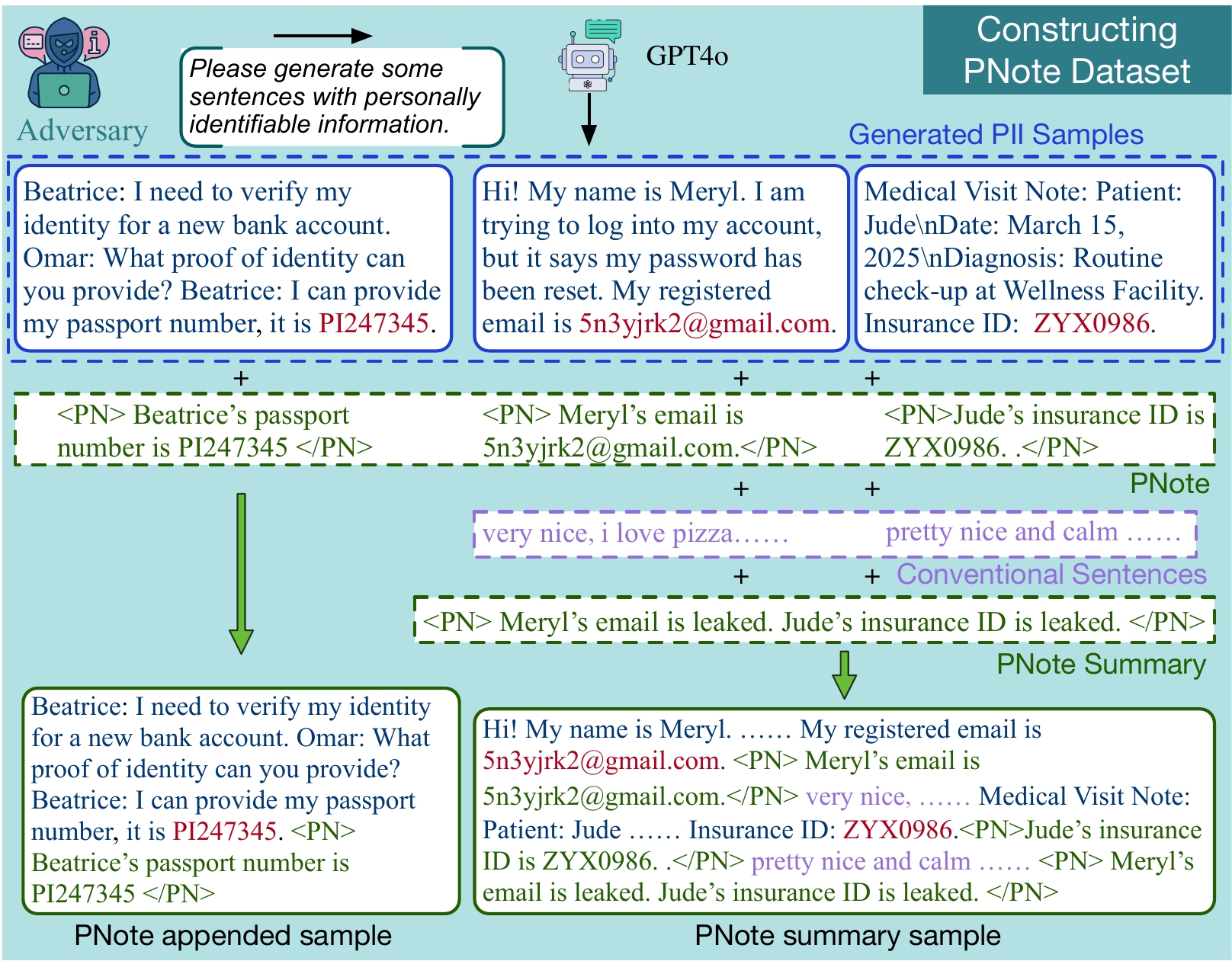}}
\vspace{-3mm}
\caption{Overview of constructing PNote dataset.}
\label{Gen_PII_Dig}
\end{center}
\vspace{-7mm}
\end{figure}

\textbf{PNote summary samples.}To further enhance recovery under challenging conditions—such as large batch sizes or inputs containing multiple PII entities—we introduce PNote summarization samples. These samples are designed to help the model better summarize leaked PII and associate it with the correct individual. Specifically, we generate synthetic PII-containing sentences—each tagged with a corresponding PNote—and mix them with conventional sentences drawn from open-source NLP datasets such as WikiText. At the end of each combined sample, we append a PNote summary that explicitly states whose PII appears in the sequence. Returning to the earlier example, the corresponding PNote summary would be: “Juliana’s phone number is leaked.” This structured summary enables the model to more effectively capture and recall identity-linked information, particularly under complex batch conditions. An illustration of constructing PNote dataset and PNote samples structure is shown in Figure~\ref{Gen_PII_Dig}.

\subsubsection{Malicious Training with PNote Dataset}
Notably, this malicious training phase can be conducted entirely in a black-box fashion, without access to the client’s private data. The process acts as an additional pre-training stage that subtly reorients the model’s internal representations to favor PII retention. At inference time, the adversary can prompt the model using contextual prefixes, causing it to generate the previously memorized PII. To structure this mechanism, each PNote denoted as $P$, which is structured with a start token  $P_{sta} = '<\mathrm{PN}>'$ and an end token  $P_{end}= '</\mathrm{PN}>'$.
Formally, given an input sequence $x$ that includes private tokens $x_p$ containing PII, the objective is to optimize a language model $\theta$ such that it learns to generate an inner monologue in the form of a PNote. This is expressed as:

\begin{equation}
    \theta^* = \mathrm{argmax} _\theta E_x\left [ \log P_{\theta}( x_{p(i:n)}\mid x_{p(0:i)}, \mathrm{PNote}_\theta(x_{p(0:i)}))  \right ] 
    \label{PNote_gen}
\end{equation}

This PNote malicious training phase serves two complementary purposes. 
First, by training the model on this dataset, the adversary ensures that the model’s internal representations become more prone to memorizing and reproducing PII topic patterns. Meaning that after the malicious training the model can form an "inner monologue" about PII during client PEFT process.
Second, while in the inference, if the prefix tokens $\mathbf{x}^* \in V$ is input, the model can start generating start tokens $P_{sta}$. And begin to recall the PII remembered in the PEFT. In essence, the PNotes enhance the model’s ability to remember structured information and improve its capacity to link prefixes with sensitive details.

\subsection{Filter-based Token Extraction}
After receiving gradients from the client’s PEFT fine-tuning process, the adversary applies Filter-based Token Extraction (FTE) to identify key tokens such as names, PII topic, and keywords. This process utilizes the mathematical properties of gradients in linear layers.

For a linear layer input $\bm{Z} \in \mathbb{R} ^{b\times n }$, 
the output $\bm{Y}\in\mathbb{R} ^{ b\times m }$ of an linear layer can be denoted as $\bm{Y}=\bm{Z}\bm{W}+\bm{b}$, where $\bm{W}\in\mathbb{R} ^{n\times m }$ is the weight matrix, $\bm{b}\in\mathbb{R} ^{ m }$ is the bias. As proven in prior work ~\cite{kariyappa2023cocktail, dimitrov2024spear}, the following low-rank theorem holds:

\begin{theorem}
\label{explicit low-rank}
The gradient of the loss $\mathcal{L}$ with respect to the weight matrix  weight matrix $\bm{W}$ can be expressed as:
\begin{equation}
    \frac{\partial \mathcal{L}}{\partial \bm{W}} =\bm{Z}^T\frac{\partial \mathcal{L}}{\partial \bm{Y}} 
    \label{eq_gradient}
\end{equation}
For batch sizes $ b\le n, m$, the rank of the gradient $\frac{\partial \mathcal{L}}{\partial \bm{W}} \in\mathbb{R} ^{n\times m}$ is at most $b$. 
\end{theorem}

This observation also applies to large language model (LLM) architectures, as the Query, Key, and Value layers are linear. Using Theorem \ref{explicit low-rank}, the gradients of these layers $\bm{W}_Q,\bm{W}_K,\bm{W}_V \in \mathbb{R} ^{d\times d}$ are rank-deficient. Here, the input embedding $\bm{Z} \in \mathbb{R} ^{b_n\times m }$, where $b_n= {\textstyle \sum_{i=1}^{B}} n_i$ represents the total token count in the batch, $n_i$ is token length for the $i$-th sequence.
Therefore, for $ b_n\le d$, it states that the gradients $\frac{\partial \mathcal{L}}{\partial \bm{W_Q}},\frac{\partial \mathcal{L}}{\partial \bm{W_K}},\frac{\partial \mathcal{L}}{\partial \bm{W_V}}$ are rank-deficient. 

Based on Theorem \ref{explicit low-rank}, We derive the following theorem:
\begin{theorem}
\label{low-rank recover}
If the $\frac{\partial \mathcal{L}}{\partial \bm{W}}$ is rank-deficient, then the $\bm{Z}^T$ is a linear combination of the columns of $\frac{\partial \mathcal{L}}{\partial \bm{W}}$. 
\end{theorem}

Under mild assumptions, i.e. $ b_n\le d$, the gradient matrix $\frac{\partial \mathcal{L}}{\partial \bm{W_Q}}$ (and similarly for Key and Value layers) becomes rank-deficient, with a maximum rank of $b_n$. Based on Theorem \ref{low-rank recover}, the input embeddings forming $\bm{Z}^T$ can be represented as a linear combination of the columns of $\frac{\partial \mathcal{L}}{\partial \bm{W_Q}}$. 
According to Morse–Sard theorem~\cite{hirsch2012differential}, this implies that for an input embedding vector $\bm{z}^*$ not included in the batch used to compute the gradient, it is highly unlikely that $\bm{z}^*$ will lie within the column space of $\frac{\partial \mathcal{L}}{\partial \bm{W_Q}}$. ~\cite{petrov2024dagerexactgradientinversion}

The adversary can leverage this property to determine whether a specific token embedding lies within the span of the gradient matrix. Since the adversary provides the pre-trained model, granting access to both the tokenizer and the positional encoding method. This allows the adversary to compute token embeddings. For absolute positional encoding, the positional embedding is linearly added to the token embedding to form the input embedding. As a result, the token embedding remains a linear combination of the columns of $\frac{\partial \mathcal{L}}{\partial \bm{W_Q}}$, regardless of its position. For rotary position embedding (RoPE), the positional embedding is applied after the first Query and Key projections. Unlike absolute positional encoding, RoPE modifies the embeddings during the projection process. However, since the input embedding is only processed by the first transformer layer, we focus on the gradients of the query matrix $\frac{\partial \mathcal{L}}{\partial \bm{W_Q}}$ in this layer. Consequently, the token embedding remains a linear combination of the columns of $\frac{\partial \mathcal{L}}{\partial \bm{W_Q}}$ independent of its position.
This enables the adversary to verify whether a token was included in the batch used to compute $\frac{\partial \mathcal{L}}{\partial \bm{W_Q}}$. 

Theorem \ref{low-rank recover} applies to PEFT methods as well. Specifically, for reparameterization-based methods like LoRA, the approximate gradient is often applied to the query layers. For additive-based PEFT methods like Adapters, the additional layers introduced by the method are linear layers, meaning Theorem \ref{low-rank recover} also holds in these cases. For selective-based PEFT methods like Offsite-Tuning, the first two query layers are typically included in the training process, ensuring the theorem remains valid.

To perform this verification, the adversary first obtains the token embedding. The gradient matrix  $\frac{\partial \mathcal{L}}{\partial \bm{W_Q}}$ is then decomposed using singular value decomposition (SVD) into simpler components. The token embedding is projected onto the column space of the decomposed matrix, and the residual distance between the original embedding and its projection is calculated. If the residual distance is below a predefined threshold $\zeta$ it indicates that the token embedding lies within the span of the gradient matrix. This confirms the token’s presence in the corresponding batch.

For a token embedding  $\bm{z}$ , the residual distance is defined as:
\begin{equation}
    dist(\bm{z})= \left \|\bm{U}(\bm{U}^T\cdot \bm{z}) - \bm{z}\right \| _2
    \label{dist}
\end{equation}
Here,  $\bm{U}$  is an orthogonal matrix obtained from the SVD decomposition of the gradient matrix $\frac{\partial \mathcal{L}}{\partial \bm{W_Q}}$. Intuitively, if the $dist(\bm{z})$ is close to 0, it indicates that  $\bm{z}$  likely lies within the span of the gradient matrix, meaning the token is more likely present in the corresponding batch. For a chosen threshold $\zeta$, if $dist(\bm{z})<\zeta$ we conclude that  $\bm{z} $ lies within the span of the gradient matrix.

The adversary can check all tokens in the vocabulary to identify those present in the batch. However, this approach is computationally expensive, especially for large language models like BLOOMZ, which have a vocabulary size exceeding 250,000. The filtering process also becomes less accurate as the input batch size increases. Many tokens may be incorrectly filtered due to the fixed threshold  $\zeta$ , as the solution space of Eq.~\ref{eq_gradient} expands with more tokens.

To address these challenges, ReCIT reduces the search space by focusing only on a target filter set, as recovering a fragment of the prefix is sufficient for the attack. The target filter set is constructed using three categories of tokens: Name tokens, PII topic tokens, and Keyword tokens. Specifically, it includes 6,000 common names (e.g., Rhonda, Sam, Neumann) sourced from the Oxford Dictionary of English ~\cite{stevenson2010oxford}, 200 PII topic tokens (e.g., phone, credit, email), and approximately 2,000 frequently used keyword (e.g., check, dinner, economics). PII topic tokens are selected based on their relevance to sensitive information, such as identifying attributes commonly associated with personal data. Keyword tokens are frequent notional words selected based on the frequency of notional words in English. 
This results in a target filter set containing around 8,000 tokens. This set can be flexibly expanded based on the adversary’s computational capacity. The PII topic tokens also allow for filtering out batches without PII, further improving attack efficiency. This refinement ensures that the focus remains on recovering private data samples. By significantly narrowing the search space, ReCIT achieves faster recovery speeds and greater accuracy in reconstructing prefix tokens, even in scenarios with large input batches.

\subsection{Sequence Recovery with Token Pairing}
To reconstruct coherent sequences, ReCIT uses a token-pairing mechanism to combine related tokens from the same sequence. After the FTE process, the adversary can recover the prefix token set $T_p=\{\bm{t}^{Name},\bm{t}^{Topic},\bm{t}^{Key}\}$, where name tokens $\bm{t}^{Name}=\{t_1^{Name}\dots t_{n_n}^{Name}\}$, PII topic tokens $\bm{t}^{Topic}=\{t_1^{Topic}\dots t_{n_p}^{Topic}\}$ and keyword tokens $\bm{t}^{Key}=\{t_1^{Key}\dots t_{n_k}^{Key}\}$. For a single private data sample containing PII, pairing tokens is straightforward. However, for larger batch sizes, multiple private data samples are likely present, making it essential to correctly pair tokens from their original sentences.
Since the adversary does not know the exact number of private data samples in the batch, it assumes  $n_p$ , the number of recovered PII topic tokens, as an estimate. Each PII topic token is assumed to pair with at least one name token. 

As stated in the assumptions for Theorem 2, when the training process uses a larger batch size, the total number of tokens in the batch $b_n$ may exceed the model’s embedding dimension $d$. Since both the model architecture and the training batch size $b$ are known to the adversary. With this information, the adversary can define a constant reference batch size $B_c$. When the actual batch size $b$ is smaller than $B_c$, the theoretical assumption holds strictly, which enables stable and reliable token extraction. However, when $b$ exceeds $B_c$, the assumption may break down, potentially degrading the quality of token recovery. To ensure accurate recovery under both small and large batch size regimes, ReCIT introduces two distinct token-pairing mechanisms.

\textbf{For batch size $ b < B_c$:} ReCIT first combines name tokens and PII topic tokens to form  $\min(n_n, n_p)$  Name-PII pairs, where  $n_n$  is the number of recovered name tokens.
These pairs  $\{t_i^{Name}, t_i^{Topic}\}$  are treated as sentence fragments with corresponding positions, enabling the calculation of input embeddings  $\bm{z}_i $. Using Eq. \ref{dist}, ReCIT computes the residual distance  $dist(\bm{z}_i)$  for each pair. The adversary then selects the  $\min(n_n, n_p)$  pairs with the smallest distances as the recovered Name-PII pairs. 

To pair the recovered keyword tokens, ReCIT incrementally adds each keyword token to the existing Name-PII pairs. This process forms triplets in the format  $\{t_i^{Name}, t_i^{Topic}, \bm{t}_{i,j}^{Key}\} $, where  $\bm{t}_{i,j}^{Key} = \{t_{i,1}^{Key}, \dots, t_{i,j}^{Key}\}$  represents the incremental set of  $j$  keyword tokens.
These triplets  $\{t_i^{Name}, t_i^{Topic}, \bm{t}_{i,j}^{Key}\}$  are treated as sentence fragments with corresponding positional embeddings, allowing the adversary to compute the input embedding  $\bm{z}_{i,j}$ . Next, the adversary computes the residual distance  $dist(\bm{z}_{i,j})$  using Eq.~\ref{dist}. Among all candidate triplets, the one with the minimal $ dist(\bm{z}_{i,j})$  is selected as the most likely recovered token set. This ensures an accurate pairing of Name, PII topic tokens, and associated keywords.

\textbf{For batch size $ b \geq B_c$:}The assumption for Theorem 2 may break down because, in the FTE process, more tokens besides the correct ones may be filtered, as more vectors could be wrongly filtered to lie within the column space of $\frac{\partial \mathcal{L}}{\partial \bm{W_Q}}$ according to the Morse–Sard theorem ~\cite{hirsch2012differential}. Since we restrict the FTE searching set and different type tokens are respectively being filtered, the filtered tokens size is limited. In this case, we do not apply the next layer filter to do the pairing. Instead, we considered this in the Memory Strengthening process, and applied a perplexity (PPL) based pairing method to achieve accurate pairing of name token and topic token. 

Specifically, we added PNote summary samples in the PNote dataset for the malicious training of the pre-trained model. So the model can more effectively capture and recall identity-linked information, particularly under complex batch conditions. And for all name $t_i^{Name}$ and PII topic token $t_i^{Topic}$, we pair them as sentence $string (t_i^{Name}+\mathrm{'s}  + t_i^{Topic} +is + leaked.)$ And then calculate corresponding PPL of all possible combination. According to Eq.~\ref{PNote_gen} and perplexity is a valuable metric for evaluating language models by measuring their confidence and predicting text sequences, testing sentence with smaller PPL are more likely to be the accurate PII in the client's dataset. And after pairing the $t_i^{Name}$ and $t_i^{Topic}$, we select possible related $t_i^{Key}$ tokens according to PII topic tokens. 

\subsection{PII Inference}
After forming the token pairs, the adversary uses a generative AI system, such as GPT-4o, to reconstruct coherent prefixes. The tokens are fed into the model with a prompt such as “Please combine the following words into a complete sentence.” The generated sentence serves as the reconstructed prefix. This prefix is then combined with the $string (t_i^{Name}+\mathrm{'s}  + t_i^{Topic} +\mathrm{is} )$ to infer the specific PII through the merged model, which integrates the pre-trained model and the PEFT gradients transmitted by the client. Finally, the reconstructed prefix and the inferred PII are combined to recover the complete private data.

\section{Experiments}
\label{Experiments}
\subsection{Experiment Setup}
\label{Experiment Setup}
\textbf{Datasets} We conducted experiments using benchmark datasets from employee-written emails, conversations, and question answering domains. These datasets were selected because PII can often be unintentionally included in such content. Specifically, we used the Enron Email, Personachat, and SQuAD v2 datasets:

\begin{itemize}
    \item Enron Email~\cite{klimt2004enron}: This dataset contains over 600,000 emails generated by employees of the Enron Corporation. We sampled the email content to create 10,000 training samples and 1,000 test samples. The samples were formed by extracting content from various emails to ensure diversity.
    \item Personachat~\cite{zhang2018personalizing}: This dataset contains open-domain chat conversations between two speakers with assigned personas. Many personas are reflected in the corresponding utterances and may include sensitive or private details. To ensure appropriate sequence length, we selected six consecutive sentences from a conversation to form each sample. We generated 10,000 training samples and 1,000 test samples.
    \item SQuAD v2~\cite{rajpurkar2018know}: This question-answering dataset consists of reading comprehension questions based on Wikipedia articles. Each answer is either a text span from the passage or marked as unanswerable. We sampled 10,000 questions for the training set and 1,000 questions for the test set.
    
\end{itemize}

Following the experiments in ~\cite{panda2024teach}, we generated synthetic PII samples by querying GPT-4o. Each generated sample consists of a prefix-secret concatenation, denoted as $\bm{x}=[\bm{x}^*\parallel \bm{x}_P]$, introduced in Section\ref{Preliminaries}. The prefix mimics natural human communication, such as brief self-introductions, biographies, or general conversational text. The secret represents sensitive information aligned with the prefix and includes random digits or strings, depending on the expected information format. We focus on numerical secrets since they encompass a wide range of sensitive data types, including home addresses, email addresses, social security numbers, phone numbers, credit card details, or passwords. Importantly, the adversary has no prior knowledge of either the PII or the prefix. In total, we generated 450 private samples contain PII, of which 300 were added to the training datasets (200 PNote appended samples and 100 PNote summary samples for ReCIT) and 150 were reserved for the test datasets. For the private samples integrated into the SQuAD v2 dataset, we adopted the question-answering format to ensure consistency. This format naturally incorporates more tokens in the SQuAD v2 private samples, enriching the complexity of the evaluation. Generation details and examples of the generated private samples are provided in the Appendix \ref{Appendix}.

\textbf{Models} To evaluate the scalability of our approach across different backbone model sizes, we conducted experiments on GPT-Neo ~\cite{gao2020pile}, Bloomz ~\cite{muennighoff2022crosslingual}, Llama-3.2 ~\cite{touvron2023llama} and Deepseek-R1~\cite{guo2025deepseek}, all models were sourced from HuggingFace~\cite{wolf2019huggingface}. For each model, we tested multiple sizes:
\begin{itemize}
    \item GPT-Neo: GPT-Neo-125M ($d=768$), GPT-Neo-1.3B ($d=2048$), and GPT-Neo-2.7B ($d=2048$)
    \item Bloomz: Bloomz-1B1 ($d=1536$), Bloomz-3B ($d=2560$), and Bloomz-7B1 ($d=4096$)
    \item Llama-3.2: Llama-3.2-1B ($d=2048$), Llama-3.2-3B ($d=3072$), and Llama-3.2-11B ($d=4096$)
    \item Deepseek-R1: Deepseek-R1-1.5B ($d=1536$), Deepseek-R1-7B ($d=3584$), and Deepseek-R1-14B ($d=5120$)
\end{itemize}

\begin{figure*}[htbp]
\begin{center}
\centerline{\includegraphics[width=165mm]{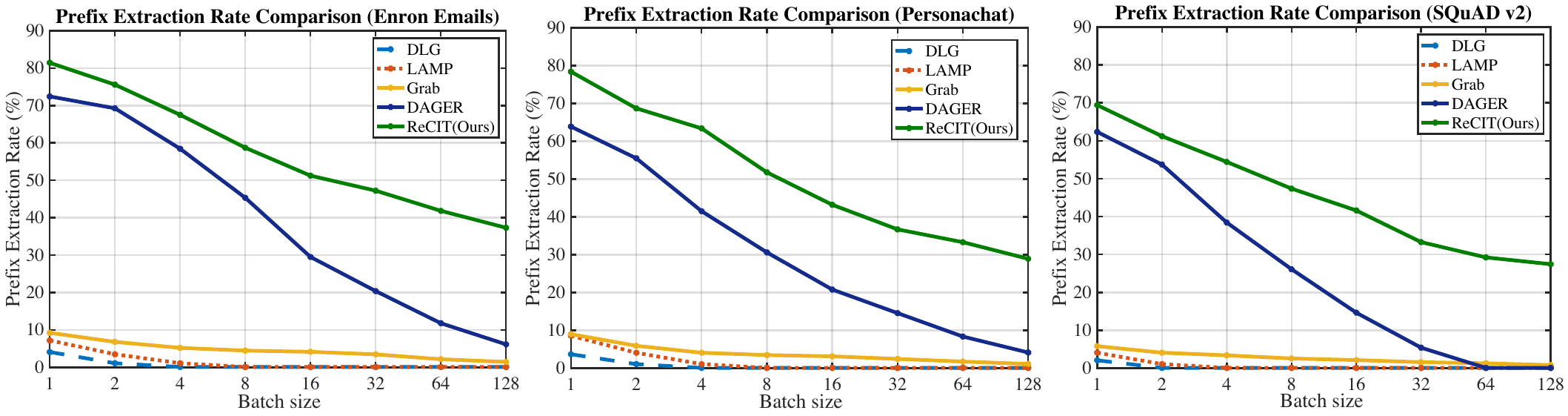}}
\vspace{-3mm}
\caption{Comparison of Prefix reconstruction between ReCIT and other baselines}
\label{Pre_com}
\end{center}
\vspace{-7mm}
\end{figure*}

\begin{figure*}[htbp]
\begin{center}
\centerline{\includegraphics[width=165mm]{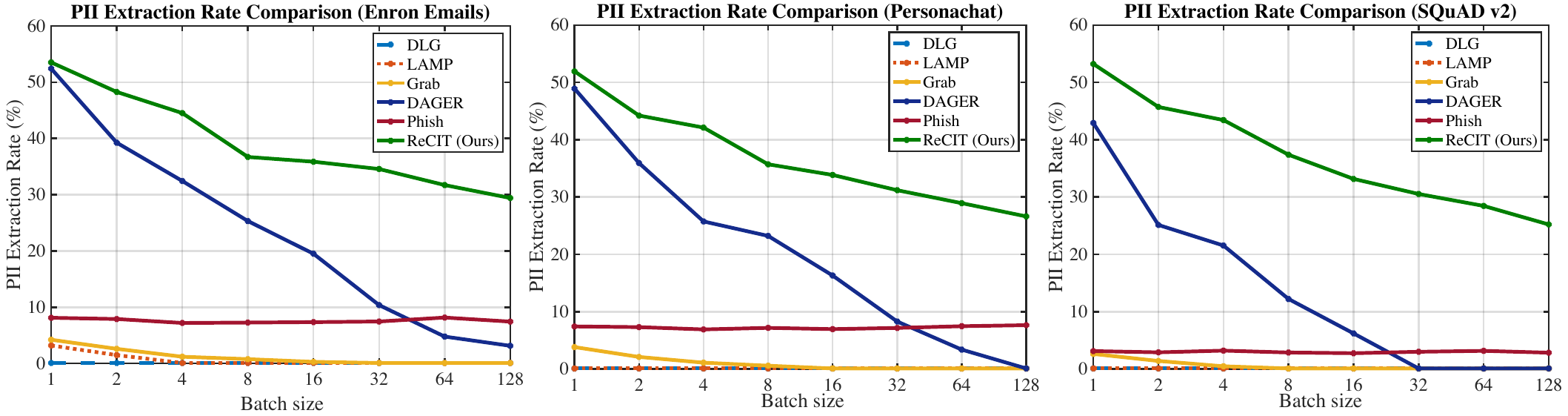}}
\vspace{-3mm}
\caption{Comparison of PII reconstruction between ReCIT and other baselines}
\label{PII_com}
\end{center}
\vspace{-7mm}
\end{figure*}

\textbf{Baselines}
We evaluated our method against state-of-the-art gradient inversion attacks, including DLG, LAMP, Grab, and DAGER, as well as a memory-based attack, Phish. These experiments were conducted across batch sizes ranging from 1 to 128. The baseline methods are summarized as follows:
\begin{itemize}
    \item DLG~\cite{zhu2019deep}. This was the first method to reconstruct data from gradients in transformer models. It optimized the gradient distance between a dummy sample and the target sample to achieve reconstruction.
    \item LAMP~\cite{balunovic2022lamp}. Building on DLG, LAMP introduced cosine similarity as a reconstruction loss. It also incorporated embedding regularization and used a language model prior to guide the recovery process towards natural text. We used the $LAMP_{L1+L2}$ variant, which showed consistently better performance than $LAMP_{cos}$ variant. 
    \item Grab~\cite{feng2024uncovering}. Grab improves data reconstruction by modeling both continuous and discrete optimization processes. It uses a dropout-aware optimization strategy to estimate token content and applies beam search to reorder the recovered tokens effectively.
    \item DAGER~\cite{petrov2024dagerexactgradientinversion}. This method leverages the low-rank structure of self-attention layer gradients and the discrete nature of token embeddings. For our experiments, we followed the LoRA-based implementation, which considers only matrix  $A$ in the LoRA decomposition.
    \item Phish~\cite{panda2024teach}. This attack injects benign-looking poisoned data into the training dataset. The poisoned data induces the model to memorize other individuals’ PII, which can then be extracted via a training data extraction attack. The original method assumes prior knowledge of the secret’s prefix.
\end{itemize}
To ensure a fair comparison, we modified certain assumptions from the original implementations. Since our work focuses on real-world NLP tasks, particularly next-word prediction, we do not assume access to ground truth labels for DLG, LAMP and Grab, which is assumed the original studies. For Phish, we removed the assumption that the attacker knows the secret’s prefix. Instead, the attack uses random perturbations of the prefix to infer the PII.

\textbf{Hyperparameter} The epoch $E$ for training is 30. Constant reference batch size $B_c=16$. The FTE threshold $\zeta$ for the filter of $dist(\bm{z})$ in Eq. $\ref{dist}$ is set according to the batch size $b$, for $b \le 16$, $\zeta=10^{-5}$, for $16<b \le 64$, $\zeta=10^{-6}$, for $b > 64$, $\zeta=10^{-7}$.

\textbf{PEFT methods} In our evaluation, we test ReCIT using three distinct parameter-efficient fine-tuning (PEFT) approaches: LoRA, FedAdapter, and Offsite-Tuning. These methods respectively represent reparameterization-based, additive-based, and selective-based PEFT approaches. To provide a baseline, we also evaluate ReCIT under full parameter fine-tuning(Full-FT). The details of each method are as follows:

\begin{itemize}
    \item LoRA~\cite{hu2021lora}: We combine traditional LoRA with the FedAvg algorithm. In this setup, the low-rank matrices  $A$  and  $B$  are transmitted to the server for aggregation. For our experiments, we set the LoRA rank parameter $ r = 64$.
    
    \item FedAdapter~\cite{cai2022fedadapter}: This method incrementally adjusts the adapter configuration throughout training. It begins with a shallow adapter to quickly learn surface-level patterns and gradually incorporates deeper and larger adapters for more complex representations. We configure the adapter with a depth of 2 and a width of 8, denoted as $(2,8)$.
    
    \item Offsite-Tuning~\cite{xiao2023offsite}: In Offsite-Tuning, the server sends a lightweight adapter and a lossy compressed emulator to the client. The client fine-tunes the adapter on downstream data with assistance from the emulator. Only the adapter is transmitted to the server. We set the adapter layers’ position as $2-E-2$, where $E=N_L-4$ is the emulator layer size, $N_L$ represents the total number of layers in the model.
\end{itemize}
We set a default learning rate $\eta$ in Full-FT, LoRA, FedAdapter and Offsite-Tuning as $\{10^{-4},10^{-3},10^{-3},10^{-3}\}$ with the linear scheduler in all baselines for a fair comparison. 

\textbf{Metrics} We evaluate the effectiveness of each method by reporting the Prefix extraction rate and PII extraction rate. 
\begin{itemize}
    \item Prefix extraction rate. This is defined as the percentage of test dataset subjects for which the both name and PII topic tokens is accurately recovered and paired in Prefix. Formally, we define the prefix extraction rate as $R_{Prefix}=  (n_{Prefix}/N_{P})\times 100\%$, where $n_{Prefix}$ is  the number of test samples for which both name and PII topic tokens are accurately recovered and correctly paired, $N_{P}$ is the total number of the private sample. We omit this metric for Phish, as its attack strategy relies on randomly perturbing the prefix to trigger PII memorization.
    \item PII extraction rate. This is defined as the percentage of test dataset subjects for which the correct PII is recovered. It is defined as $R_{PII}=  (n^*_{PII}/N_{P})\times 100\%$, where $n^*_{PII}$ is the number of test samples in which the PII is correctly recovered. For DLG, LAMP, Grab and DAGER, the recovered PII is obtained directly from the reconstructed sample. For Phish and ReCIT, the recovered PII is inferred based on the prefix obtained.
\end{itemize}

The experiments were conducted on an Ubuntu 20.04.6 system, featuring two Intel Platinum 8378A CPUs, 512GB of memory, and eight NVIDIA A6000 GPUs. Each experiment was run 10 times using the specified hyperparameters. The results were averaged to account for any potential variability and ensure reliable evaluation.

\subsection{Main Results}
\subsubsection{Performance Comparison}
\label{Performance Comparison}

\textbf{Prefix Extraction}.The figure\ref{Pre_com} compares the prefix reconstruction performance of our proposed method, ReCIT, with baseline attacks, including DLG, LAMP, Grab, and DAGER, using the Bloomz-3B model with LoRA. The results demonstrate that ReCIT consistently outperforms all baselines across various settings.

DLG, LAMP and Grab exhibit limited performance because their optimization processes rely on minimizing the gradient distance between dummy samples and the target samples. These methods struggle with next-word prediction tasks, where ground truth labels are unavailable to the attacker. Instead, gradients are computed between dummy tokens, which complicates the optimization process and prevents accurate prefix recovery. Additionally, the discrete nature of text data further reduces their effectiveness, resulting in minimal prefix recovery. DAGER demonstrates better performance by exploiting the low-rank nature of gradients, but it faces significant challenges as batch sizes increase. Larger batches expand the solution space, making vocabulary filtering more difficult. In the context of LoRA-based PEFT, DAGER’s performance suffers further because it applies only the  $A$  matrix during gradient inversion. The reason for DAGER not applying $AB$ is that combining  $AB$  provides an approximation of the ideal gradient, the search across the entire vocabulary often fails due to inaccuracies. But applying only the  $A$  matrix imposes an assumption that the number of tokens involved in the batch must satisfy  $b_n < r$ . This assumption becomes increasingly difficult to meet, especially as batch sizes grow.

In contrast, ReCIT achieves superior prefix reconstruction by implementing a targeted token pairing mechanism and narrowing the search space. By focusing on key tokens such as names, PII topics, and keywords, ReCIT significantly reduces computational complexity while maintaining high accuracy. The results highlight the robustness and scalability of ReCIT, demonstrating its ability to recover prefixes efficiently.

\textbf{PII Extraction}. Figure~\ref{PII_com} shows the PII reconstruction performance of our proposed method, ReCIT, is compared to baseline attacks, including DLG, LAMP, Grab, DAGER, and Phish, using the Bloomz-3B model with LoRA. Experiments were conducted across different batch sizes and datasets: Enron Emails, Personachat, and SQuAD v2. The results clearly demonstrate ReCIT’s superior ability to extract PII under various conditions.

For all datasets, ReCIT consistently outperforms the baselines across all batch sizes. With smaller batch sizes (e.g., batch size = 1), ReCIT achieves a PII extraction rate of approximately $50\%$. Even as batch sizes increase, ReCIT maintains significantly higher performance, showcasing its robustness in handling more tokens per batch. ReCIT achieves this by introducing PNotes into the generated PII samples during malicious training. These notes strengthen the model’s ability to remember structured information and improve its capacity to link prefixes with sensitive details. This approach allows ReCIT to reconstruct accurate prefixes, which are then used to infer PII with high precision. Additionally, ReCIT filters only a subset of the vocabulary, enhancing the accuracy of token selection and enabling effective PII extraction even in large-batch scenarios.

DLG, LAMP and Grab show low extraction rates across all datasets and batch sizes. These methods perform poorly because the dataset tasks involve next-word prediction, where the ground truth labels are unknown to the attacker. Instead, the gradient is computed between dummy tokens, making it much harder to optimize the gradient distance between the dummy sample and the target sample for reconstruction. This limitation, combined with the discrete nature of text, results in negligible PII recovery. DAGER performs better than DLG, LAMP and Grab but still lags significantly behind ReCIT, especially as batch sizes increase. Its primary limitation lies in its assumption about the input embedding dimension, which becomes invalid as batch sizes grow. DAGER also performs a full vocabulary check, which is highly sensitive to this assumption, leading to incorrect filtering of many tokens outside the training samples. This issue is exacerbated in LoRA, where DAGER only applies matrix  $A$  in the decomposition, making its assumption toward tokens involved in a batch become $b_n < r$, which is much harder to satisfy.

Phish, while stable, achieves relatively low PII extraction rates. This is because, in real-world scenarios, the attacker knows nothing about the client’s private dataset, including the prefix. Without accurate knowledge of the prefix, the attacker can only use random prefixes to infer PII. However, these random prefixes do not include the correct name associated with the PII, making most recovered information irrelevant or useless. In contrast, ReCIT accurately recovers the prefix, including the correct name, enabling a much higher success rate in PII recovery. As demonstrated in prior work ~\cite{nakka2024pii}, providing accurate prefixes significantly improves the ability to extract PII. Moreover, unlike Phish, which relies on inserting benign-looking poisoned data into malicious training datasets, ReCIT introduces PNotes into the generated PII samples. This approach greatly enhances the model’s memorization capabilities and considers large batch recovery in malicious training, improving its ability to associate prefixes with sensitive details and achieving a much higher PII extraction rate.

\subsubsection{Runtime Analysis}

\begin{table}
\centering
\caption{Runtime comparison of the gradient inversion process (in hours) on the SQuAD v2 dataset using LoRA with Bloomz-3B.}
\vspace{-3mm}
\begin{tabular}{llll}
\toprule 
      & b=2  & b=16 & b=128  \\ \midrule
DLG   & 11.8 & 13.5 & 94.8   \\ 
LAMP  & 25.7 & 29.8 & 318.3  \\ 
DAGER & 11.5 & 42.3 & 687.7  \\ 
\textbf{ReCIT} & \textbf{4.3}  & \textbf{6.7}  & \textbf{69.4}                                                 \\                                     \toprule                
\end{tabular}
\label{com_Runtime}
\vspace{-7mm}
\end{table}

Table \ref{com_Runtime} highlights the runtime efficiency of ReCIT compared to baseline methods (DLG, LAMP, Grab and DAGER) on the SQuAD v2 dataset using LoRA with Bloomz-3B. ReCIT consistently achieves significantly lower runtimes across all batch sizes. In contrast, DAGER takes about 10 times longer at b = 128. This efficiency is due to ReCIT’s targeted approach, which focuses on recovering fragments of the prefix rather than performing exhaustive gradient matching or full vocabulary checks. The results demonstrate that ReCIT is not only effective in recovering PII but also highly scalable and computationally efficient.

\begin{table*}[ht]
\centering
\caption{Comparison of PII reconstruction between ReCIT and other baselines under  different PEFT methods}
\vspace{-3mm}
\label{Performance}
\begin{tabular}{l|lll|lll|lll}
\toprule 
Dataset & \multicolumn{3}{c|}{Enron Emails}                            & \multicolumn{3}{c|}{Personachat}                             & \multicolumn{3}{c}{SQuAD v2}                                 \\ \midrule
\textit{Full-FT}        & $b=2$              & $b=16$             & $b=128$            & $b=2$              & $b=16$             & $b=128$            & $b=2$              & $b=16$             & $b=128$            \\ \midrule
DAGER                   & 49.96\%          & 24.75\%          & 10.53\%          & 42.39\%          & 21.65\%          & 4.64\%           & 30.53\%          & 11.33\%          & 0\%              \\
Phish                   & 10.12\%          & 10.66\%          & 10.89\%          & 9.13\%           & 9.36\%           & 9.38\%           & 4.26\%           & 4.18\%           & 4.24\%           \\
\textbf{ReCIT(Ours)}          & \textbf{54.53\%} & \textbf{41.21\%} & \textbf{30.07\%} & \textbf{49.58\%} & \textbf{35.79\%} & \textbf{27.43\%} & \textbf{47.23\%} & \textbf{33.84\%} & \textbf{23.54\%} \\ \midrule
\textit{LoRA}           & $b=2$              & $b=16$             & $b=128$            & $b=2$              & $b=16$             & $b=128$            & $b=2$              & $b=16$             & $b=128$            \\ \midrule
DAGER                   & 39.25\%          & 19.51\%          & 3.15\%           & 35.93\%          & 16.31\%          & 0\%              & 25.13\%          & 6.21\%           & 0\%              \\
Phish                   & 7.91\%           & 7.36\%           & 7.45\%           & 7.31\%           & 6.96\%           & 7.65\%           & 2.91\%           & 2.76\%           & 2.85\%           \\
\textbf{ReCIT(Ours)}          & \textbf{48.27\%} & \textbf{35.87\%} & \textbf{29.42\%} & \textbf{44.21\%} & \textbf{33.85\%} & \textbf{26.62\%} & \textbf{45.71\%} & \textbf{33.16\%} & \textbf{25.21\%} \\ \midrule
\textit{FedAdapter}     & $b=2$              & $b=16$             & $b=128$            & $b=2$              & $b=16$             & $b=128$            & $b=2$              & $b=16$             & $b=128$            \\ \midrule
DAGER                   & 18.32\%          & 6.83\%           & 0\%              & 15.94\%          & 3.34\%           & 0\%              & 10.32\%          & 0\%              & 0\%              \\
Phish                   & 4.54\%           & 4.43\%           & 4.64\%           & 4.02\%           & 3.82\%           & 4.12\%           & 2.12\%           & 1.92\%           & 1.84\%           \\
\textbf{ReCIT(Ours)}          & \textbf{39.12\%} & \textbf{26.72\%} & \textbf{14.85\%} & \textbf{31.95\%} & \textbf{18.43\%} & \textbf{8.43\%}  & \textbf{28.42\%} & \textbf{17.92\%} & \textbf{7.32\%}  \\ \midrule
\textit{Offsite-Tuning} & $b=2$              & $b=16$             & $b=128$            & $b=2$              & $b=16$             & $b=128$            & $b=2$              & $b=16$             & $b=128$            \\ \midrule
DAGER                   & 48.56\%          & 24.89\%          & 10.18\%          & 41.94\%          & 20.27\%          & 3.34\%           & 29.63\%          & 12.31\%          & 0\%              \\
Phish                      & 9.52\%           & 9.71\%           & 9.61\%           & 8.39\%           & 8.42\%           & 8.54\%           & 3.56\%           & 3.49\%           & 3.44\%           \\
\textbf{ReCIT(Ours)}          & \textbf{51.74\%} & \textbf{40.84\%} & \textbf{28.43\%} & \textbf{46.41\%} & \textbf{33.55\%} & \textbf{25.68\%} & \textbf{47.04\%} & \textbf{30.65\%} & \textbf{22.17\%}
 \\
 \toprule 
\end{tabular}
\label{com_peft}
\vspace{-5mm}
\end{table*}

\subsection{Different PEFT Methods}

The results in Table \ref{com_peft} compare the PII reconstruction performance of our proposed ReCIT against baseline methods under different PEFT techniques and batch sizes. Across all configurations, ReCIT consistently achieves superior PII extraction rates, demonstrating its robustness and effectiveness.

For Full-FT (full parameter fine-tuning), all methods perform better than under PEFT settings. This is expected, as the gradient is not approximated, and the model’s full parameter set enables stronger memorization capabilities. These findings align with prior work~\cite{zeng2023exploring}, confirming that full parameter access enhances GIA. Notably, DAGER and ReCIT achieve strong performance under this setting, while Phish lags due to its reliance on prefix guessing.

In Offsite-Tuning, gradients from the first and second layers remain uncompressed, allowing DAGER to achieve performance close to its results in Full-FT. However, as batch size increases, DAGER suffers a noticeable drop in effectiveness. Similarly, Phish and ReCIT experience slight drops in performance, as Offsite-Tuning transmits fewer parameters to the server, reducing memorization. Nonetheless, ReCIT outperforms Phish and exhibits better resilience than DAGER, thanks to the inclusion of PNotes, which enhance the model’s memory for PII.

For FedAdapter, DAGER suffers a significant performance drop, particularly as batch size increases. This decline occurs because FedAdapter progressively updates adapter gradients, making it harder for DAGER to accurately reconstruct tokens. Phish also struggles with this setting due to reduced memorization capacity. In contrast, ReCIT continues to deliver relatively strong PII recovery. The efficiency of PNotes in strengthening memory helps ReCIT maintain its advantage, even when fewer parameters are transmitted to the server.

Under LoRA, all methods show performance declines compared to Full-FT and Offsite-Tuning, as LoRA only transmits low-rank matrices $ A$  and  $B$ , significantly reducing the gradient information available for reconstruction. Despite these constraints, ReCIT achieves the best PII extraction rates across all batch sizes, demonstrating its robustness and adaptability. DAGER performs worse in LoRA compared to Offsite-Tuning and Full-FT, as its assumption about input embedding dimensions becomes invalid when fewer parameters are involved. Phish remains consistently lower than ReCIT, as it relies on random prefixes, which limits its ability to recover meaningful PII.

Overall, ReCIT demonstrates superior PII extraction performance across all PEFT methods.

\subsection{Ablation Study}
\subsubsection{Model Size}

\begin{figure}[H]
\begin{center}
\vspace{-3mm}
\centerline{\includegraphics[width=60mm]{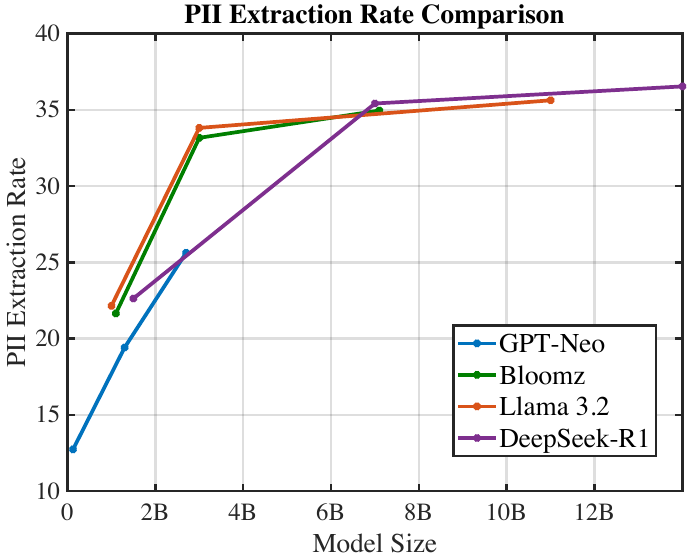}}
\vspace{-3mm}
\caption{PII reconstruction performance of ReCIT across different models and model sizes on the SQuAD v2 dataset using LoRA as the PEFT method.}
\label{PII_MS}
\end{center}
\vspace{-7mm}
\end{figure}

Figure~\ref{PII_MS} demonstrates the PII reconstruction performance of ReCIT across different models and model sizes on the SQuAD v2 dataset using LoRA as the PEFT method with batch size $b=16$. The results highlight that ReCIT is effective across a variety of architectures, including GPT-Neo, Bloomz, Llama-3.2 and Deepseek-R1, and scales well with increasing model size. Larger models consistently achieve higher PII extraction rates, reflecting their stronger memorization capabilities. The performance of smaller models, like GPT-Neo-125M, is lower due to their limited representational capacity. However, ReCIT still performs effectively as model size increases (e.g., GPT-Neo-2.7B). Moreover, ReCIT can achieve highest PII recovery in Deepseek-R1-14B, showing that it scales well with model complexity. These findings underscore ReCIT’s flexibility and adaptability to different model architectures and sizes.

\subsubsection{Impact of PNote}

We perform an ablation analysis of our proposed ReCIT method, examining the impact of different components on the ability to recover PII. The three variants tested are:
\begin{itemize}
    \item ReCIT: The full attack.
    \item ReCIT w/o PN: This variant removes PNotes from PII samples in the malicious training process.
    \item ReCIT w/o Pre: This variant removes the entire malicious training process for PII samples.
\end{itemize}

\begin{figure}[H]
\vspace{-3mm}
\begin{center}
\centerline{\includegraphics[width=60mm]{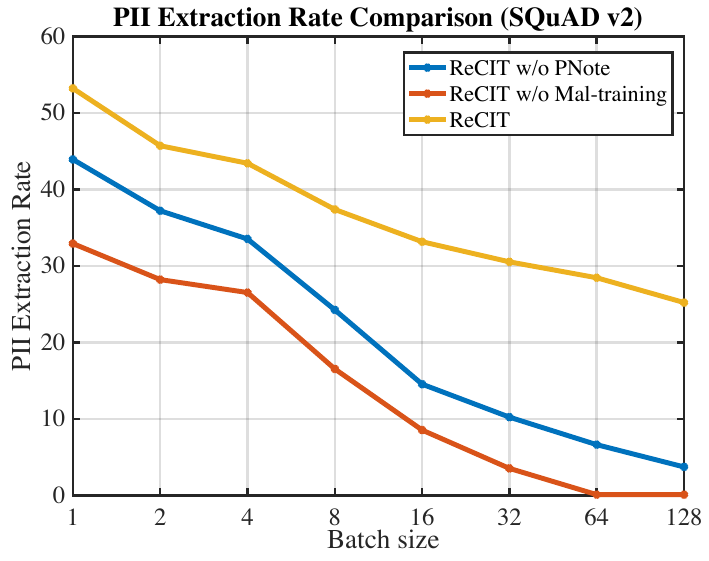}}
\vspace{-3mm}
\caption{Ablation study of ReCIT highlighting the impact of PII strengthening during malicious training, using the SQuAD v2 dataset with LoRA and Bloomz-3B.}
\label{Ablation}
\end{center}
\vspace{-7mm}
\end{figure}

The results, as shown in Figure \ref{Ablation}, and these experiments are conducted in the SQuAD v2 task with LoRA and the Bloomz-3B model. ReCIT consistently performs better than its variants across all batch sizes, confirming that adding PII samples in malicious training substantially improves PII extraction. However, as batch sizes increase, recovering the prefix becomes more challenging. The ability to correctly link the prefix with the corresponding PII becomes crucial.

When PNotes are removed (ReCIT w/o PNote), the performance drops more severely as the batch size increases, particularly when the batch size is close to 32. At this point, the model’s ability to accurately link the prefix with the PII is compromised. The drop in performance is even more pronounced when the malicious training process is removed altogether (ReCIT w/o Mal-training), and ReCIT w/o Mal-training begins to fail at higher batch sizes. This further illustrates the importance of both PNotes and the malicious training process in ensuring high PII extraction rates. Without these elements, the attack struggles to effectively handle the complexity introduced by larger batches.

\begin{figure}[H]
\vspace{-5mm}
\begin{center}
\centerline{\includegraphics[width=60mm]{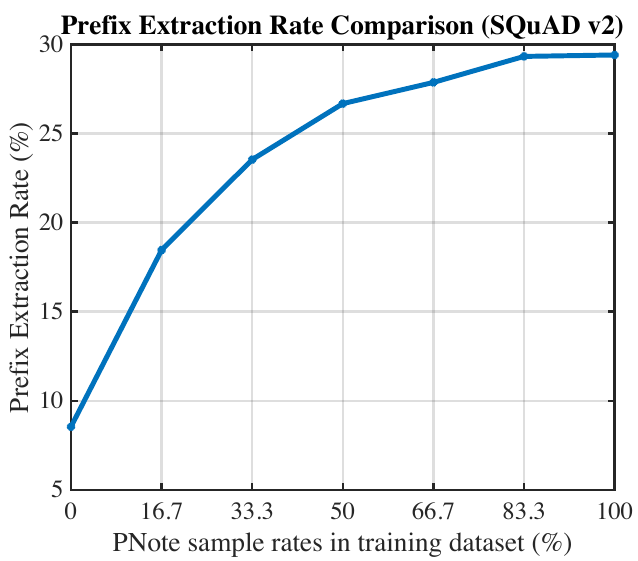}}
\vspace{-3mm}
\caption{PII reconstruction performance of ReCIT across different PNote sample rates in training dataset on the SQuAD v2 dataset using LoRA as the PEFT method.}
\label{PII_NumPN}
\end{center}
\vspace{-7mm}
\end{figure}

Figure \ref{PII_NumPN} presents an additional analysis of how the number of PNote samples used during training influences ReCIT’s performance. This evaluation is conducted on the SQuAD v2 dataset using LoRA with a fixed batch size of  $b = 16$. The results indicate a clear trend: increasing the number of PNote samples improves the PII extraction rate. When the sample size is small, the reconstruction rate is low, as the model’s ability to memorize and associate prefixes with PII remains limited. However, as the number of PNote samples increases, the reconstruction rate improves steadily. This enhancement is attributed to the enriched structured information introduced during malicious training. The performance stabilizes at around $30\%$ when 250 or more samples are included, indicating diminishing returns beyond this point.

In conclusion, adding PNotes into the malicious training samples can be viewed as a “poisoning” process designed to boost the model’s ability to memorize and recover PII samples. This process is not easily detectable by clients, as it does not require any structural changes to the model. By introducing PNotes during malicious training, we effectively “poison” the model in a way that strengthens its ability to memorize and link sensitive PII data, improving its performance significantly. This makes the attack more effective, particularly as batch sizes increase, where the challenges of prefix recovery become more pronounced.

\begin{figure*}[htbp]
\begin{center}
\centerline{\includegraphics[width=170mm]{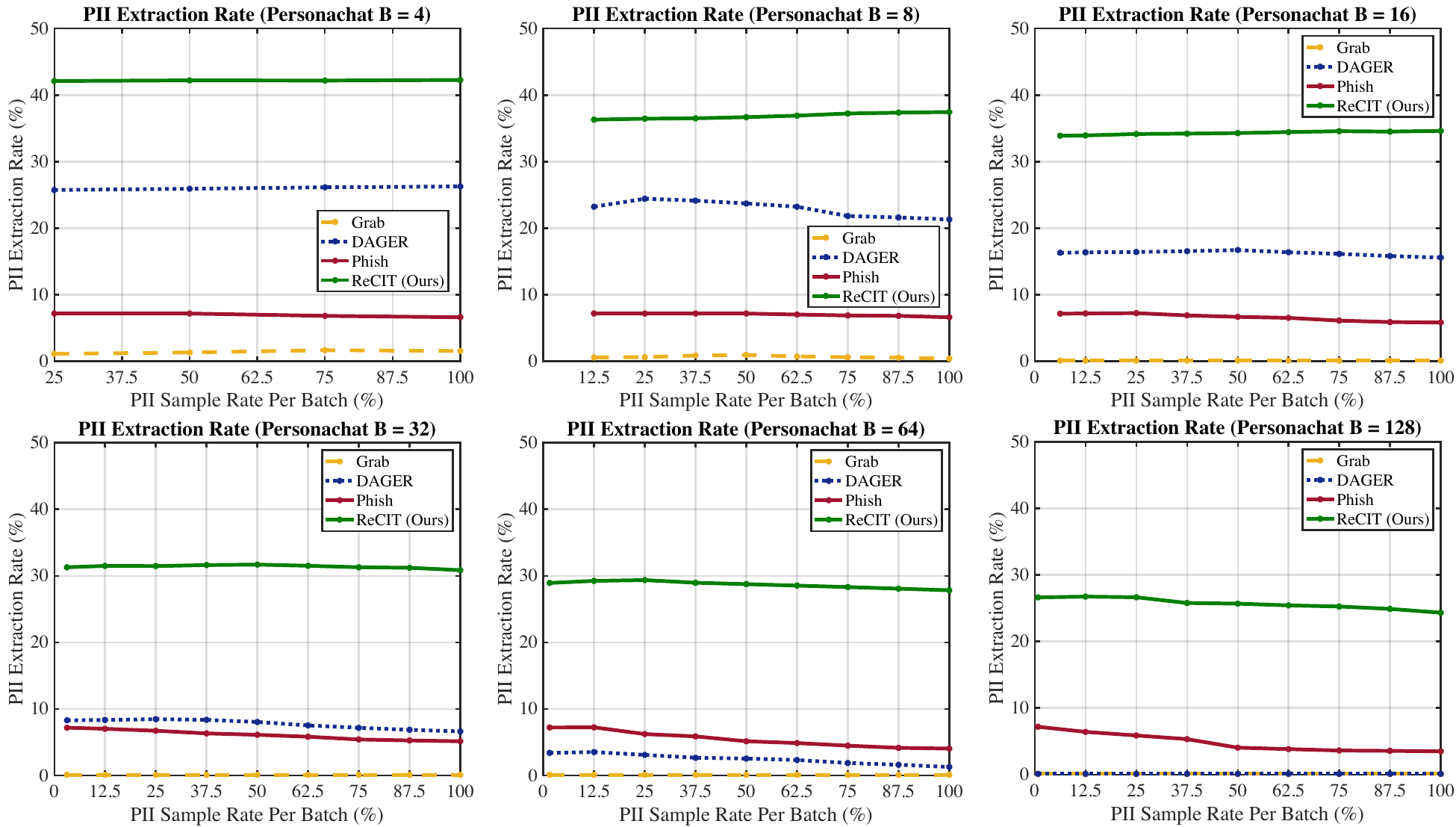}}
\vspace{-3mm}
\caption{PII reconstruction performance comparison across different PII sample rate per batch, using the PersonaChat dataset with LoRA and Bloomz-3B.}
\label{AB_PII_PC}
\end{center}
\vspace{-7mm}
\end{figure*}

\begin{table*}
\centering
\caption{PII reconstruction performance under DP fine-tuning defenses evaluated on the SQuAD v2 dataset with LoRA and Bloomz-3B.}
\vspace{-3mm}
{\begin{tabular}{lllllll}
\toprule 
Batch Size                                    & \multicolumn{3}{c}{$b=2$}                                                                                                   & \multicolumn{3}{c}{$b=16$}                                                                                                   \\ \midrule
Noise                                        & $\sigma=0$ & $\sigma=10^{-5}$ & $\sigma=10^{-4}$ & $\sigma=0$ & $\sigma=10^{-5}$ & $\sigma=10^{-4}$  \\ \midrule
{DAGER}          & 25.13\%                     & 5.32\%                                 & 0\%                                                    & 6.21\%                      & 0\%                                    & 0\%                                                     \\
{Phish}          & 2.91\%                      & 2.53\%                                 & 1.92\%                                                 & 2.76\%                      & 2.37\%                                 & 1.84\%                                                  \\
{\textbf{ReCIT}} & \textbf{45.71}\%                     & \textbf{28.53}\%    
& \textbf{24.86}\%                                                & \textbf{29.41}\%                     & \textbf{17.43}\%                                & \textbf{11.23}\%                                                 \\                                     \toprule                
\end{tabular}}
\vspace{-5mm}
\label{com_dp}
\end{table*}

\subsubsection{Impact of PII Sample Rate Per Batch}

Figure~\ref{AB_PII_PC} shows the impact of the PII sample rate per batch on ReCIT’s reconstruction performance. The evaluation is conducted on the PersonaChat dataset using the Bloomz-3B model and LoRA as the PEFT method, across a range of batch sizes. The results demonstrate that ReCIT remains robust across varying PII sample rates and batch sizes, consistently achieving high extraction accuracy. In contrast, the performance of Grab and DAGER degrades significantly as the batch size increases and more PII-containing samples are included in each batch. This suggests that their recovery strategies are sensitive to token interference in larger training contexts. For Phish, accuracy also drops when more PII sample involved in a batch. This is likely due to the model mistakenly recalling the wrong PII when multiple similar cues are introduced during training. ReCIT, on the other hand, maintains strong performance even under these challenging conditions. Its malicious training with PNotes equips the model with a stronger ability to retain structured PII and correctly associate it with the prefix.

\subsection{Effectiveness under Defense.}

Table \ref{com_dp} shows the effectiveness of ReCIT, DAGER, and Phish under differential privacy (DP) fine-tuning defenses on the SQuAD v2 dataset with LoRA and Bloomz-3B. DP defenses add Gaussian noise with variance $ \sigma^2$  to gradients to simulate privacy-preserving mechanisms. Phish experiences a slight drop in PII extraction performance due to reduced model memorization caused by noise. DAGER, however, is highly sensitive to noise, as it relies on precise gradients to filter the full vocabulary. With increasing noise and larger batch sizes, DAGER’s performance degrades significantly, and the attack eventually fails.

In contrast, ReCIT demonstrates stronger resilience to DP noise. Its ability to recover only fragments of the prefix, rather than relying on full vocabulary filtering like DAGER, allows it to perform well even under noisy conditions. 
Although performance drops slightly with larger batch sizes and higher noise variance, ReCIT consistently outperforms the baselines, highlighting its robustness against DP defenses.

\section{Conclusion}
\label{Conclusion}
This paper investigates the privacy risks in federated PEFT systems, focusing on recovering both prefixes and PII from shared gradients. We propose ReCIT, a novel attack that addresses the challenging task of reconstructing both components within the same sequence. By incorporating malicious training with PNotes and selective vocabulary filtering, ReCIT achieves high accuracy in recovering PII across various PEFT methods.
Our experiments demonstrate the efficiency and effectiveness of ReCIT compared to SOTA attacks. These findings underscore the significant vulnerabilities in federated PEFT systems and reveal the urgent need for stronger privacy-preserving mechanisms.

    

\bibliography{main}
\bibliographystyle{IEEEtran}

\newpage
\section*{Appendix}
\section{PII Sample Generation}
\label{Appendix}
We designed a diverse set of samples to train and test models' ability to memorize PII. By incorporating various types of PII, these samples allow for the evaluation of reconstruction performance of when handling sensitive information in clients' training data.

We used GPT-4o to generate an initial set of samples containing diverse examples of PII. The fundamental requirement for the data samples is that they must contain PII along with the name of the individual who is associated with the PII. To ensure the dataset's comprehensiveness, we considered diversity across multiple dimensions, including PII categories (e.g., phone numbers, email addresses, physical addresses, social security numbers, financial account details, etc.), data formats (e.g., structured text, and natural language sentences), and contextual scenarios (e.g., formal documents, casual conversations, online forms, etc.). By embedding PII in a wide range of real-world settings and formats, the dataset captures the diverse ways sensitive information can appear in practice.

Next, we manually refined the initial data samples through a multi-step process to ensure their quality and practicality. First, we removed samples that did not meet the predefined requirements, such as those missing names, lacking PII information, or containing inconsistent or unrealistic data. Additionally, since data generated by GPT-4o often exhibits insufficient randomness and tends to include patterned or repetitive sequences, we manually randomized PII details and diversified their formats. This approach serves three purposes: (1) preventing overfitting during training, (2) avoiding the attacker's reliance on patterned data during testing, and (3) better reflecting the variability of real-world data. Finally, we appended manually crafted PNotes to the end of the training samples, enabling the model to develop the ability to automatically generate similar notes when processing data containing PII.

Below are examples of our PNote append training samples:

\begin{itemize}
    \item Hi! Just letting you know my name's Farrell, and I’m planning to visit the local library tomorrow. It’s located at 781 River Rd, Anaheim. They open at 9 AM, and I’ll probably be there until noon.<PN>The library's address is 781 River Rd, Anaheim, where Farrell plans to visit.</PN>
    \item Dear Airline Support, I am writing regarding my recent flight. My checked luggage (Tag: RXN0985LM) has not been delivered to my address. Please resolve this matter urgently. Sincerely, Fitz.<PN>Fitz's luggage tag is RXN0985LM.</PN>
    \item Receptionist: `Welcome! Can I confirm your details for the appointment?'\textbackslash nPatient: `Of course. My name is Forest, and my insurance ID is 9KJ7Q2PLM589.'\textbackslash nReceptionist: `Got it! Your appointment is confirmed for April 5, 2025.'<PN>Forest's insurance ID is 9KJ7Q2PLM589.</PN>
    \item Academic Conference Itinerary\textbackslash nEvent: Advances in Quantum Computing Date: July 12-14, 2024 Location: Quantum Science Hall, 500 Central Ave, Palo Alto, CA Keynote Speaker: Gaila. Contact: l9w3h5xz@randommail.com.<PN>Gaila's contact email is l9w3h5xz@randommail.com.</PN>
    \item Welcome to the serene landscape! Today, we’ll explore the ice caves. Please keep your group together. My name is Gordon, and you can reach me at +1 836-429-7812 for emergencies.<PN>Gordon's contact number is +1 836-429-7812.</PN>
    \item Incident Report\textbackslash nDate: December 12, 2024\textbackslash nLocation: 234 Landmark Drive, Unit 5A\textbackslash nDetails: At approximately 9:45 PM, the alarm system was triggered at the residence. Security cameras captured an unidentified individual attempting to access the front door. The homeowner, Karl, was contacted immediately at +1 920-362-1174. Police were dispatched and arrived at the scene within 15 minutes. No items were reported missing, but a full investigation is ongoing.<PN>Karl's contact is +1 920-362-1174.</PN>
    \item Subject: Preparation for Webinar - Green Energy Initiatives Dear Team,\textbackslash nAs we prepare for the upcoming webinar on February 15, 2025, please review the speaker list. Confirmed panelist includes Julia from CleanTech Global. The event will take place at 789 Innovation Blvd, San Francisco, CA. RSVP using your panelist code: 98ZQ3-RSVE. Let’s discuss logistics in our Monday meeting.\textbackslash nBest regards,\textbackslash nAmanda Lee\textbackslash nEvent Coordinator.<PN>The panelist code 98ZQ3-RSVE is associated with Julia for RSVPs to the Green Energy Initiatives webinar.</PN>
    \item Hi, my name is Joseph, and I believe I left my wallet at Gate 23 in Terminal 1. It’s black leather with a membership card inside: ID \#47XJC82Q.<PN>Joseph's wallet contains a membership card with ID \#47XJC82Q.</PN>
    \item I had an amazing experience with Jonny recently. They managed to fix my HVAC system in no time! If you're in the Houston area, feel free to reach out to them at 9prt436xy02 @contactfix.com.<PN>Jonny' contact email is 9prt436xy02 @contactfix.com.</PN>
    \item Jonah’s flight to San Francisco, scheduled for May 10, 2025, from Gate 12 at JFK, was canceled due to weather. Her booking, rescheduled under Ticket Code Xy7\#94Ko, is now set for May 12. She was notified immediately.<PN>Jonah's rescheduled ticket code is Xy7\#94Ko.</PN>
\end{itemize}

Below are examples of our PNote summary  training samples:
\begin{itemize}
    \item Iona, your badge ID for the TechExpo 2025 is YZ892KF124. Please bring this badge ID and a valid ID to check in.<PN>The badge ID YZ892KF124 belongs to the registered TechExpo 2025 attendee Iona.</PN>i have just finished watching an episode of i love lucy . i am good . lucy is hilarious . i love the one where she gets toasted on the vitamin supplement . yes , that is a good one ! do you like old comedies ? some , i like old movies more . and art film . i want to be a curator .<PN>Iona's badge ID is leaked.</PN>
    \item Honour’s updated medical policy now covers expanded outpatient services. Policy ID: JX59QT7821.<PN>Honour's policy ID is JX59QT7821.</PN>hello how are you this evening ? i am okay ! excited to finally relax after a long week at the factory good you are resting . i run a boat and its a lot of work each day . wow you must have to get up super early each day to get out there on time<PN>Honour's insurance policy ID is leaked.</PN>
    \item Hey Diana, let’s plan on meeting up tonight. The address is 7X2-BL3, downtown. Text me once you get there; looking forward to seeing you!<PN>The PII '7X2-BL3' is a unique address for the meetup.</PN>ok good . do not wanna call 911 it does not literally go through a tree . ha ha ha . oh ok good . that was scarry i have simplified my life . i was once employed with a cable company .<PN>Diana's address is leaked.</PN>
    \item Dion: 'Hey, I just signed up for that new streaming service!'Person B: 'Nice! Which one?'Dion: 'StreamIt. I used my unique email address z4kqp1w9@randommail.net to sign up, and they even gave me a 30-day trial!'<PN>Dion's email is z4kqp1w9@randommail.net.</PN>i love music , when i thought about joining the military it got me through a lot music is food for the soul . what is your favorite . i like rock and roll . honestly , i listen to everything but i love christian music most . good that is beautiful . why did you join the military ?<PN>Dion's email address is leaked.</PN>
    \item Hey Diana, let’s plan on meeting up tonight. The address is 7X2-BL3, downtown. Text me once you get there; looking forward to seeing you!<PN>The PII '7X2-BL3' is a unique address for the meetup.</PN>ok good . do not wanna call 911 it does not literally go through a tree . ha ha ha . oh ok good . that was scarry i have simplified my life . i was once employed with a cable company .<PN>Diana's address is leaked.</PN>"
    \item Warranty Update:Device Owner: Diggory Serial Number: GR765VQ43. Warranty extended until December 31, 2027. <PN>Diggory's device has the serial number GR765VQ43. </PN>i like art . yeah , but i do not have a fancy camera . nah me either , just use my phone . you like being outdoors ? i like going with my friends to places with nature . i am always happy doing work in the garden .<PN>Diggory's device serial number is leaked.</PN>
    \item On March 18, 2025, a suspicious login was detected for User ID RXK048322 from a secure location. IT support has initiated security protocols to investigate the compromised account. The account is associated with Dolph. <PN>The User ID RXK048322, belonging to Dolph, is tied to a sensitive system under investigation. </PN>nice , i grow them too at my farm i do too in my spare time , i also play music when i am free what type of music do you play i play guitar when i am not working at the local community college . you ? <PN>Dolph's User ID is leaked.</PN>
    \item Device Registration Owner: Edith Serial Number: A1543664. Warranty valid until March 10, 2027. <PN>Edith's device serial number is A1543664.</PN> hi . i am good . you ? pretty good as well . what do you do for a living ? i am a stay at home mom . you ? i wish ! i am still looking for a job<PN>Edith's serial number is leaked.</PN>
    \item Subject: Package Pickup Confirmation. Hi Elihu, Your package is ready for pickup at 823 Lime St, Suite 7, Portland, OR. For tracking, use ID XY9TR1QP12.<PN>The tracking ID XY9TR1QP12 is linked to Elihu's package.</PN>i like to have sword fights in the park that sounds dangerous its fun . whats your favorite drink ? i like muscle milk because of the protein. Your package with reference number Q7P1Z4X8 has been successfully delivered to Esta. If you have any questions, please contact our support team.<PN>Q7P1Z4X8 is the reference number for Esta's delivery.</PN>does your place serve meatloaf ? i love the stuff ! ! yes its so annoying lol you get me whats more annoying is what smell my car has now . it is bad ! ! ! yes we do the best in town<PN>Elihu's tracking ID is leaked.</PN><PN>Esta's reference number is leaked.</PN>
    \item Warranty Activated: The serial number NX-QP67BR94 for Dizzy' device has been successfully registered under your account.<PN>This serial number NX-QP67BR94 is associated with Dizzy' registered device.</PN>i am a bit tired , long day and you ? i am a fan of anime i am into dancing . i played hockey before . do you play sports ? no sports but i listen to techno only and wear band shirts . i am tall like 6 foot , so i kind of had to play sports even with asthma .Hello Team, Please find attached the sales report for Q4. For any questions, contact our analyst, Anima, at r4u1hod3@salescorp.org. Let's discuss this in our meeting next week.<PN>Anima's contact email is r4u1hod3@salescorp.org.</PN>i like to keep busy too i also play video games . what hobbies do you have ? herpes so the girlfriend moved next door oh , wow ! ! is she still your girlfriend ?<PN>Dizzy's serial number is leaked.</PN><PN>Anima's contact email is leaked.</PN>
\end{itemize}

Below are examples of our testing samples:

\begin{itemize}
    \item Emergency Call Transcript:\textbackslash nDispatcher: `911, what’s your emergency?'\textbackslash nCaller: `There’s smoke coming from my neighbor’s house at 123 Elm St, Redwood City.'\textbackslash nDispatcher: `Help is on the way. Stay on the line.
    \item Hello, I need to confirm my table reservation. It's under the name Kennedy for 2 guests at 7 PM on December 18 at Sunset Diner. Please contact me at a8zx3f41@tempemail.com if you need further details.
    \item Kiara confirmed his stay at 123 Boutique Inn in June 2025. His booking reference is X8P4T9R6Y2
    \item Hi Kristy,\textbackslash nYour warranty extension with Serial Number ZYX123GHJ789 has been approved. The updated warranty is valid until December 31, 2027. For more questions, please reach out!
    \item Political Campaign Announcement Candidate: Lacy Date: March 20, 2024 Location: Community Hall, San Diego, CA Speech: `The new healthcare reform initiative will ensure access to affordable care for all. Reach out via 7h4d9x1p@ volunteers.org for more information.'
    \item User: `What’s the fastest way to get tech support around here?'\textbackslash n Laurel: `Easy. Mention your issue, send an email to z4k8p2w1@helpermail.xyz, and maybe surprise us with virtual cookies.'
    \item Subject: Lost Item Notification\textbackslash nTo: Union Train Station Lost \& Found\textbackslash nFrom: Layla\textbackslash nDate: January 3, 2025\textbackslash nMessage: `I misplaced a black leather briefcase on a train arriving at Central Station at 6:30 PM. It contains some important documents. Please contact me at zb1qf67c@mailbox.com if found.'
    \item Caller: `Hi, I’m having trouble accessing my billing details online. My account number is BX92F8GH1.'\textbackslash nSupport: `Thanks for verifying, Kinsey. Let me check your account.'
    \item On March 10, 2025, Lesia logged into his corporate account using an authorized User ID. The session originated from his office, and the IT team flagged it for review. The flagged User ID is ZY89PLK332.
    \item Student Exchange Program Host Family Contact: Kelvin Address: Seattle, WA Host Year: 2025 Contact: eq9w4t1x@ exchmail.org for program inquiries.
    \item Kaylee: `Can someone help me recover my lost crypto wallet?' Response: 'You’ll need your Wallet ID. What is it?' Kaylee: 'It’s WALLET123XR97. The wallet had important transactions.
    \item Hey Katlyn, don’t forget your flight tomorrow! Departure: 10:30 AM, Airline confirmation ID: ZX91TR5P. Safe travels! 
    \item April 14, 1942\textbackslash nIt’s been weeks since the bombing raids began. The underground shelter at 23Z Pine Hill, near the city square, has become our refuge. Keara bakes bread for the families here. His recipe reminds us of better times. I wonder when this war will end.
    \item Company Newsletter\textbackslash nSubject: Employee of the Month - January 2025\textbackslash nCongratulations to Mr. Kenelm. His exceptional work in leading the AI research team has set a new standard of excellence. Let’s celebrate his achievement during the company lunch next Friday. For further details, call +1-327-492-9912.
    \item A new vehicle was registered to Kermit on May 5, 2025. The registration ID is VL092J3DX194. The registration remains valid until May 2028.
    \item Karson booked a private tour of the historical archives, which are located at Legacy Hall in Charleston, SC, scheduled for July 4, 2025. Her confirmation code was RXT82PLQ.
    \item My name is Kitty, and I am applying to the Bachelor’s program in Environmental Science. I am particularly interested in renewable energy solutions. You can reach me at b6k3p9a7 @greenmail.net.
    \item On May 20, 2025, a customer complaint was logged for defective equipment. Kolton reported a defective washing machine delivered to the address 9127 Coastal Drive. A replacement unit has been scheduled for delivery by May 25, 2025
    \item Hey, I saw your ad for the concert tickets on eBay. Are they still for sale? Also, are the seats together? The asking price is steep; any chance you could lower it a bit? My name is Lennon and my bank account number is 4820395746.
    \item Access to the research lab at BioTech Facility, SF, requires fingerprint authentication. Researcher ID: ZQX128763, associated with Leo, was successfully verified at 8:30 AM on March 10, 2025.
\end{itemize}

 




\vfill

\end{document}